\documentclass{article}
\usepackage{spconf,amsmath,graphicx,hyperref}
\usepackage{graphicx}
\usepackage{subcaption}
\usepackage{booktabs}
\usepackage{amsfonts}
\usepackage{musicography}
\usepackage{tabularx}
\usepackage{multirow}
\usepackage{comment}

\title{Rate-Agnostic Bioacoustics: heterogeneous Multi-Taxa Classification with Continuous Filterbanks and fourier Neural Operators}
\name{Stefano Ciapponi\textsuperscript{1,2,\musSharp}%
\thanks{\musSharp\ These authors contributed equally to this work.},
Francesco Ardan Dal Rí\textsuperscript{1,\musSharp}, Nicola Conci$^{1}$, Elisabetta Farella $^{2}$}
\address{University of Trento$^{1}$, Fondazione Bruno Kessler$^{2}$}
\begin{document}
\maketitle

\vspace{-0.4cm}
\begin{abstract}
Conventional bioacoustic classification models rely on fixed-rate spectral representations, requiring recordings acquired at heterogeneous sampling rates to be resampled before analysis. We propose a Sampling-Frequency-Independent (SFI) frontend that processes each recording directly at its native sampling rate, coupled with a Fourier Neural Operator (FNO) backbone featuring progressive temporal-scale fusion. This framework avoids fixed-rate resampling and high-frequency information loss while producing fixed-size representations across sampling rates. Mild training-time sampling-rate (\textit{sr}) augmentation further improves robustness to unseen rate variations. Evaluated on a multi-taxa corpus comprising 84 classes and 60 sampling rates, the proposed SFI-FNO configuration outperforms fixed-rate and corpus-maximum-rate baselines, achieving .906 accuracy, .921 balanced accuracy, and a Macro-F1 score of .899.
\end{abstract}


\vspace{-0.2cm}
\section{Introduction}
\label{sec:intro}
\vspace{-0.1cm}
Bioacoustic signal processing underpins critical applications in ecological monitoring, biodiversity conservation, and animal behavior analysis \cite{gibb2019emerging, sugai2019terrestrial}. However, real-world deployments rely on field data collected across highly heterogeneous hardware \cite{hill2018audiomoth}, resulting in extreme sample-rate variation, ranging from $~300\text{~Hz}$ acoustic monitoring to over $250\text{~kHz}$ ultrasonic recording. Standard deep learning pipelines commonly rely on fixed, discrete matrix representations, such as Short-Time Fourier Transforms (STFT) or Mel-scale spectrograms \cite{stowell2022computational}. Because these discrete matrix structures tie coordinate resolution directly to the sampling frequency, variations in sample rate alter input matrix dimensions and feature alignment, forcing expensive digital resampling, model retraining, or multi-model switching.
Additionally, recent bioacoustic models often rely on large pretrained feature extractors, such as AVES \cite{hagiwara2023aves}, Bird-MAE \cite{rauch2025birdmae}, or Perch~2.0 \cite{van2025perch}. These models typically operate at fixed input rates between $16$\,kHz and $32$\,kHz. Although sufficient for much of the audible spectrum, these rates impose Nyquist limits that are critical for taxa whose calls extend far into the ultrasonic range \cite{aodha2022towards, hill2018audiomoth}. This motivates models that can operate directly on heterogeneous native sampling rates (\textit{sr}s) without sacrificing available spectral information, unlocking adaptive \textit{sr} changes on autonomous recording units and avoiding model finetuning.

\begin{figure}[t]
    \centering
    \includegraphics[width=0.92\linewidth]{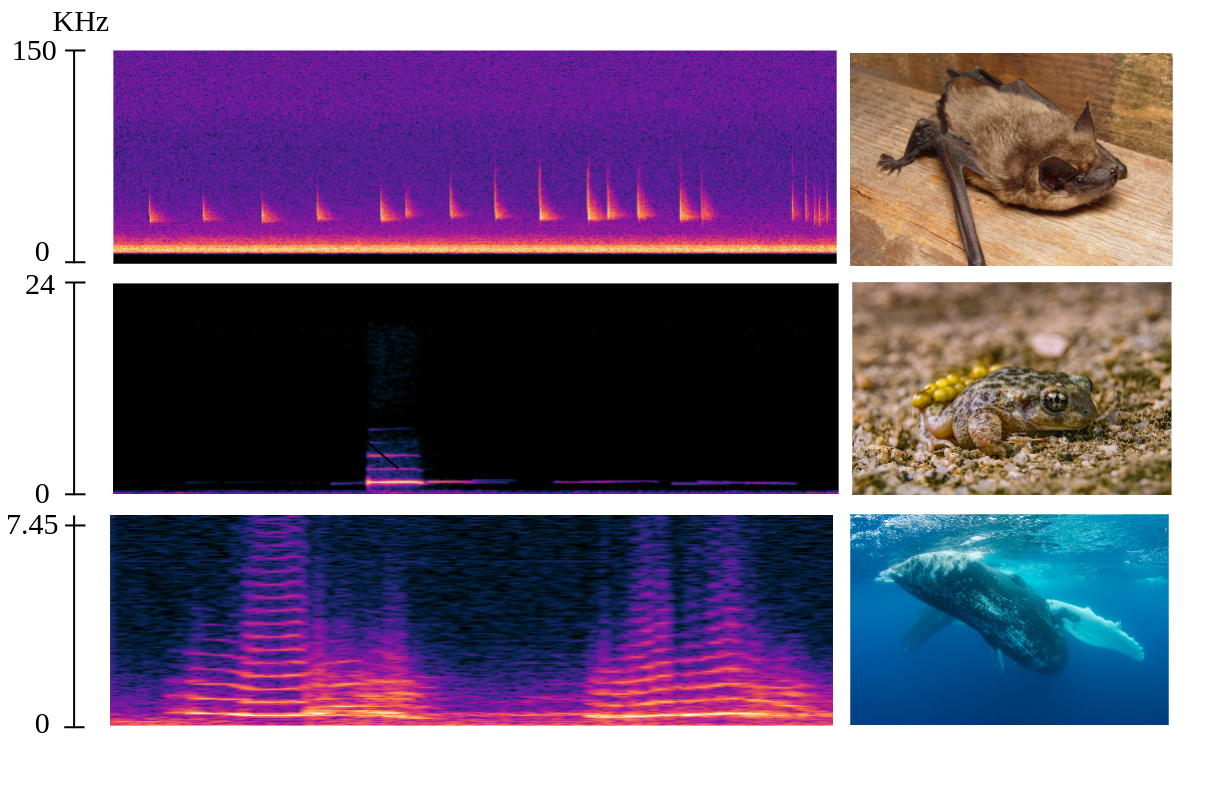}
    \vspace{-0.3cm}
    \caption{\centering Examples of different spectrograms for different taxa, emphasizing the large sample-rate variability.}
    \label{fig:intro}
    \vspace{-0.5cm}
\end{figure}

To cope with such sample-rate bottleneck, we embrace the signal-processing paradigm of ``\textit{think analog, act digital}'' \cite{ThinkAnalog}. We propose a Sampling-Frequency-Independent (SFI) continuous-frequency neural filterbank operating directly at the native input rate, coupled with a Fourier Neural Operator (FNO) \cite{li2021fourier} backbone for learning representations over the resulting log-frequency-time feature grids. In detail, first, the SFI frontend parameterizes a bank of continuous log-frequency responses in physical frequency (Hz), which are discretized according to each recording's native sampling frequency and evaluated through rate-conditioned analytic filterbanks. Nyquist-aware masking and physical-time envelope sampling provide a common feature organization without resampling recordings to a fixed target rate. Second, the FNO backbone processes the resulting multi-scale log-frequency-time representations through spectral and local transformations, followed by progressive scale fusion. We evaluate the contribution of both components through a factorial comparison against fixed-rate and corpus-maximum Mel frontends, plus CNN and MBConv backbones. Finally, we demonstrate the efficiency advantages for edge deployments, including mild \textit{sr} augmentation to suppress rate-dependent shortcuts without requiring per-rate model switching or retraining.

\section{Methodology}
\vspace{-0.1cm}
\subsection{Sampling-Frequency-Independent Frontend}
\label{sec:sfi_frontend}

Let $\mathbf{x}[n] \in \mathbb{R}$, $n = 1, \dots, N$, denote a discrete-time audio signal obtained by sampling a continuous-time source at an arbitrary native rate $f_s$. The Sampling-Frequency-Independent (SFI) frontend processes $\mathbf{x}[n]$ at its native rate $f_s$ without resampling it to a common target sampling frequency, constructing multi-scale representations across $M$ temporal scale configurations $(\tau_m, \Delta t_m) \in \mathcal{W} \times \mathcal{H}$ for $m = 1, \dots, M$, where $\mathcal{W}$ and $\mathcal{H}$ denote the sets of candidate window lengths and hop sizes, respectively.

Target center frequencies $f_k$ ($k = 1, \dots, K$) are logarithmically spaced from $f_{\min}$ to $f_{\max}$, with each band $k$ parameterized by a constant log-frequency concentration parameter $Q$. To model continuous spectral profiles, each band $k$ is parameterized by a continuous frequency-domain response $W_k(f)$ for $f \in \mathbb{R}^+$:
{\small
\begin{equation}
\begin{aligned}
    W_k(f) &= W_{k,\text{base}}(f)\,
    \exp\left[\beta R_{\theta,k}(f)\right], \\
    W_{k,\text{base}}(f) &= \exp\left(
    -\frac{1}{2}
    \left[
    Q \ln \frac{\max(f,\epsilon_f)}{f_k}
    \right]^2
    \right),
\end{aligned}
\end{equation}
}
where $W_{k,\text{base}}(f)$ is a continuous log-Gaussian bandpass profile, $\epsilon_f>0$ avoids the logarithmic singularity at zero frequency, $R_{\theta,k}(f) \in [-1,1]$ is the $k$-th output of a frequency-response residual MLP employing Fourier feature embeddings, and $\beta$ controls the maximum residual adaptation in the log-magnitude domain.

To prevent the use of center frequencies exceeding the native Nyquist limit of $x[n]$, a dynamic binary validity mask $m_k \in \{0, 1\}$ is applied:
{\small
\begin{equation}
m_k = \mathbb{I}\left(f_k \le \eta \cdot \frac{f_s}{2}\right),
\end{equation}
}
where $\eta \in (0, 1)$ represents the Nyquist roll-off factor. For computational efficiency, valid bands are assigned to an anti-aliased dyadic waveform pyramid and processed at an internal rate $f_{s,k}=f_s/2^{d_k}$ selected according to $f_k$.

To construct discrete time-domain filter taps of length $L$, the continuous responses $W_k(f)$ are evaluated at physical frequencies $f_{p,k}=p f_{s,k}/N_{\mathrm{FFT}}$. The real and quadrature filter responses are synthesized on-the-fly via inverse Fourier transforms:
\begin{equation}
\begin{aligned}
    \tilde{h}_{k,\text{real}} &= \mathcal{F}^{-1}\{W_k(f_{p,k})\}, \\
    \tilde{h}_{k,\text{imag}} &= \mathcal{F}^{-1}\{-jW_k(f_{p,k})\},
\end{aligned}
\end{equation}
with the DC and Nyquist quadrature components set to zero. The impulse responses are centered, Hann-windowed, and jointly normalized to obtain $(h_{k,\text{real}},h_{k,\text{imag}})$.

To decouple downstream temporal resolution from $f_s$, the band-pass analytic envelope is evaluated directly at a constant physical frame period $\Delta t_{\text{base}}=4.0\,\text{ms}$:
{\small
\begin{equation}
e_k[t] = m_k
\sqrt{
\big[(x^{(d_k)} * h_{k,\text{real}})[tH_k]\big]^2 +
\big[(x^{(d_k)} * h_{k,\text{imag}})[tH_k]\big]^2
},
\end{equation}
}
where $H_k=\max(1,\lfloor \Delta t_{\text{base}}f_{s,k}\rceil)$ and $x^{(d_k)}$ denotes the waveform at the corresponding dyadic level. Two representation channels ($C=2$) are computed per band: a log-compressed magnitude envelope, $\log(1+e_k[t])$, and Per-Channel Energy Normalization (PCEN) \cite{lostanlen2018per} with trainable band-wise smoothing and gain control parameters.

Finally, the base feature tensor $\mathbf{X} \in \mathbb{R}^{K \times 2 \times T_{\text{base}}}$ is downsampled to target scales $\tau_m$ via 1D average pooling with
{\small
\begin{equation}
P_m=\max\left(1,\left\lfloor\frac{\tau_m}{\Delta t_{\text{base}}}\right\rceil\right),
\qquad
R_m=\max\left(1,\left\lfloor\frac{\Delta t_m}{\Delta t_{\text{base}}}\right\rceil\right),
\end{equation}
}
producing multi-scale feature maps $\mathbf{X}_m \in \mathbb{R}^{K \times 2 \times T_m}$, which are standardized per representation channel over valid frequency--time entries.

\label{sec:complexity}
By subsampling early inputs to a fixed feature size, the framework enables downstream processing to run at constant complexity regardless of sampling rate, allowing variable inputs without model re-allocation or audio resampling.
\begin{figure}[h]
    \centering
    \includegraphics[width=0.8\linewidth]{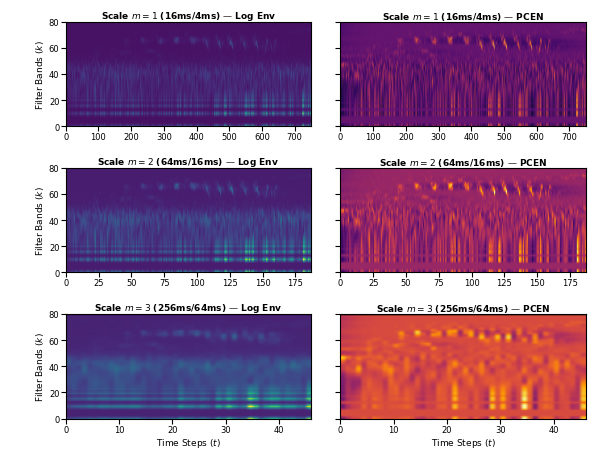}
    \vspace{-0.2cm}
    \caption{\centering SFI multi-scale features applied to a \textit{Fringilla coelebs} audio clip at native 24\,kHz using 16, 64, and 256\,ms  scales.}
    \label{fig:sfi_example}
\end{figure}

\vspace{-0.6cm}
\subsection{Fourier Neural Operator Backbone}
\vspace{-0.1cm}
To extract representations prior to classification, the feature backbone employs two-dimensional Fourier Neural Operators (2D FNO)~\cite{li2021fourier}. Rather than relying exclusively on local spatial kernels as in standard CNNs, FNO layers perform global spectral transformations over the log-frequency--time feature grid. Given a feature map $v_l$, a 2D FNO block updates the representation via:

\begin{equation}
\begin{gathered}
u_l = Wv_l + \mathcal{F}^{-1} \left\{R_\theta \odot \mathcal{F}\{v_l\}\right\}, \\
v_{l+1} = v_l + \mathcal{D}_p\!\left(\sigma\!\left(\mathrm{GN}(u_l)\right)\right),
\end{gathered}
\end{equation}

where $\mathcal{F}$ and $\mathcal{F}^{-1}$ denote the 2D FFT and its inverse across the frequency and time axes, $R_\theta$ parameterizes a truncated set of learned complex Fourier modes, $W$ is a local $1\times1$ linear transformation, $\mathrm{GN}$ denotes Group Normalization, $\sigma$ is a GELU activation, and $\mathcal{D}_p$ denotes the dropout operator with probability $p$. The resulting feature representation is subsequently aggregated through masked global pooling into a fixed-dimensional latent vector for class logit projection.

\vspace{-0.2cm}
\subsection{Scale Fusion and Latent Projection}
\label{ssec:scale_fusion}

To process the multi-scale feature maps $\mathbf{X}_m \in \mathbb{R}^{K \times 2 \times T_m}$ ($m \in \{1,2,3\}$) generated by the SFI frontend (Sec.~\ref{sec:sfi_frontend}), we employ a hierarchical \textit{Temporal Pyramid Encoder}. The frontend extracts features using matched window-hop pairs $(\tau_m,\Delta t_m) \in \{(16,4),(64,16),(256,64)\}\,\text{ms}$ - a $75\%$ temporal overlap across all scales.

Processing begins at the finest temporal scale $\mathbf{X}_1$. Each scale representation is projected to the corresponding stage dimension $C_m$ and masked using a binary spatial validity matrix $\mathbf{M}_m \in \{0,1\}^{B \times 1 \times K \times T_m}$. For subsequent scales $m>1$, intermediate hidden maps $\mathbf{H}_{m-1}$ are adaptively average-pooled along the time axis to match the native frame length $T_m$, concatenated with the projected scale features, and fused via $1\times1$ convolutions with GroupNorm and GELU. This preserves the native temporal resolution of each frontend scale without explicit interpolation.

Following the final backbone stage, the feature map $\mathbf{H} \in \mathbb{R}^{B \times C_{\text{tail}} \times K' \times T'}$ is aggregated under the projected spatial mask $\mathbf{M}$. Dual masked global pooling combines mean and peak responses:
{\small
\begin{equation}
\mathbf{p} =
\left[
\frac{\sum_{k,t}\mathbf{H}\odot\mathbf{M}}
{\max(1,\sum_{k,t}\mathbf{M})}
\;\Bigg\|\;
\max_{(k,t):M_{k,t}=1}\mathbf{H}_{k,t}
\right],
\label{eq:masked_pool}
\end{equation}
}
where $\mathbf{p} \in \mathbb{R}^{2C_{\text{tail}}}$. A linear projection with LayerNorm maps $\mathbf{p}$ to a $192$-dimensional latent embedding $\mathbf{z}$, feeding the primary linear classifier alongside a 2-layer contrastive head ($192 \to 192 \to 128$) for representation learning.

\vspace{-0.2cm}
\section{Experimental Setup}
\vspace{-0.2cm}
\subsection{Datasets}
\vspace{-0.1cm}
To evaluate performance across diverse taxa, species, and native sample rates, we assembled a composite dataset. Taken together, this corpus presents significant challenges, including severe imbalances in class distribution, sample counts, and sampling frequencies.

\begin{figure*}[t]
    \centering
    \includegraphics[width=0.8\linewidth]{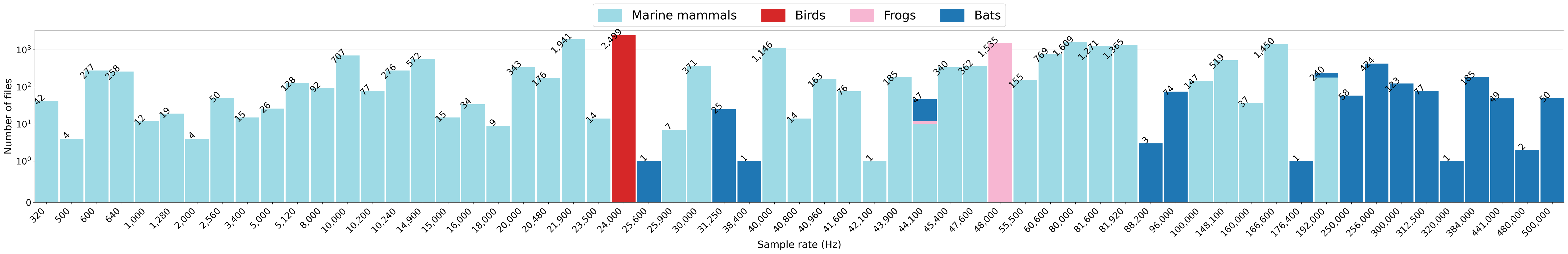}
    \vspace{-0.2cm}
    \caption{\centering Number of files in the entire corpora for every individual native sample rate (log scale for better visualization).}
    \label{fig:sample_rate_distribution}
\end{figure*}

Specifically, we collected: the Full-Cut version of the Watkins Marine Mammal Sound Database \cite{sayigh2016watkins}, a wide collection of historic recordings spanning more than 60 marine species; the BioDCASE 2026 Task 3 development set \cite{biodcase_tiny_2026_dataset}, comprising 10 European bird species; the anuran dataset by Akbal et al. \cite{akbal2023explainable}, containing 26 frog and toad species; and the Bat Calls dataset \cite{ndiaye2026bat_calls}, comprising ultrasonic recordings from 6 bat species. These corpora span markedly different taxonomic groups, spectral ranges, and class distributions, providing a heterogeneous benchmark for evaluating classification performance across different native sampling frequencies. Following a filtering criterion similar to other works, e.g., \cite{licciardi2024whalenet}, we excluded species represented by fewer than 15 samples, thus avoiding extremely underrepresented classes. 
Details and statistics over the entire corpora are summarized in Table \ref{tab:dataset_statistics}, while Figure \ref{fig:sample_rate_distribution} highlights the distribution of individual sample rates.

\begin{table}[h!]
\centering
\small
\setlength{\tabcolsep}{4pt}
\caption{Corpora statistics ($\mu/\sigma$ after class filtering).}
\vspace{-0.1cm}
\label{tab:dataset_statistics}
\begin{tabular}{lrrrr}
\toprule
Dataset & Classes & Samples/class & Rates & Rate (kHz) \\
\midrule
Marine \cite{sayigh2016watkins} & 42 & $362 \pm 485$ & 43 & $62.8 \pm 49.8$ \\
Birds \cite{biodcase_tiny_2026_dataset} & 10 & $250 \pm 0$   & 1  & $24.0 \pm 0.0$ \\
Frogs \cite{akbal2023explainable} & 26 & $59 \pm 14$   & 2  & $48.0 \pm 0.1$ \\
Bats \cite{ndiaye2026bat_calls} & 6  & $198 \pm 48$  & 19 & $273.4 \pm 108.5$ \\
\midrule
\textbf{Total} & \textbf{84} & $\mathbf{243 \pm 368}$ & \textbf{60} & $\mathbf{69.2 \pm 72.6}$ \\
\bottomrule
\end{tabular}
\end{table}

\vspace{-0.5cm}
\subsection{Model Configurations and Hyperparameters}
\label{ssec:model_configs}

To systematically evaluate the SFI frontend and FNO backbone against fixed-rate and corpus-max baselines, we construct a $3 \times 3$ experimental suite, combining three frontends with three backbone architectures. Besides SFI and FNO, we included two other multiscale Mel-spectrogram frontends, namely (i) MEL48, resampled to $48$\,kHz and (ii) MAX, resampled to the maximum corpus \textit{sr}, and two other backbones, namely (i) CNN, a 2D residual ConvNet, and (ii) MBC, an EfficientNet-style Mobile Inverted Bottleneck backbone \cite{tan2019efficientnet}. 
All configurations use $K=128$ frequency bins and the same window-hop pairs $(\tau_m,\Delta t_m)$ as in \ref{ssec:scale_fusion}. For SFI, $f_{\min}=5$\,Hz, $Q=12$, and $\eta=0.95$, with $f_{\max}=\eta f_{s,\max}/2$. The frequency-response residual MLP uses a hidden dimension of $64$, $6$ Fourier-feature bands, and residual scale $\beta=1.5$; filter kernels have length $L=129$ and are synthesized using a $512$-point FFT. PCEN uses a $64$-frame causal smoothing support. 

For the robustness ablation, SFI-FNO was additionally retrained across all five folds with random \textit{sr} augmentation (0.75--1.05$\times$, $p=.85$).

Models are trained with AdamW for up to $80$ epochs using an initial learning rate of $3\times10^{-4}$, weight decay $10^{-4}$, batch size $16$, and two-step gradient accumulation. The objective combines class-weighted focal loss \cite{lin2017focal} with supervised contrastive loss \cite{khosla2020supervised}, weighted by $0.10$; validation Macro-F1 drives scheduling and early stopping. For conciseness, additional details, results, and code will be made available at the repository: \texttt{[AVAILABLE UPON ACCEPTANCE]}.

\vspace{-0.1cm}
\section{Results}
\vspace{-0.2cm}
Experimental results, summarized in Table~\ref{tab:classification_results}, show that SFI consistently improves over MEL48 and MAX across backbones. SFI-FNO performs best overall, reaching .906 accuracy, .921 balanced accuracy, and .899 Macro-F1. Relative to MEL48, SFI improves accuracy by 5.2, 5.7, and 4.7 points for CNN, MBC, and FNO, respectively, indicating that most of the gain arises from the frontend, with a complementary benefit from FNO.


\begin{table}[htbp]
    \centering
    \footnotesize
    \setlength{\tabcolsep}{4.8pt}
    \caption{Classification results (reported as mean $\pm$ std.)}
    \label{tab:classification_results}
    \vspace{-0.1cm}
    \begin{tabular}{@{}llcccc@{}}
        \toprule
        \textbf{Front.} & \textbf{Back.} & \textbf{Params.} & \textbf{Accuracy} & \textbf{Bal. Acc.} & \textbf{Macro-F1} \\
        \midrule
        \multirow{3}{*}{MEL48}
            & CNN    & 916,496   & .831 $\pm$ .019 & .881 $\pm$ .009 & .847 $\pm$ .017 \\
            & MBC & 854,532   & .838 $\pm$ .019 & .886 $\pm$ .009 & .852 $\pm$ .016 \\
            & FNO    & 1,007,048 & .860 $\pm$ .008 & .893 $\pm$ .002 & .866 $\pm$ .004 \\
        \midrule
        \multirow{3}{*}{MAX}
            & CNN    & 916,496   & .860 $\pm$ .014 & .894 $\pm$ .004 & .865 $\pm$ .006 \\
            & MBC & 854,532   & .888 $\pm$ .008 & .914 $\pm$ .009 & .892 $\pm$ .010 \\
            & FNO    & 1,007,048 & .895 $\pm$ .010 & .910 $\pm$ .008 & .890 $\pm$ .010 \\
        \midrule
        \multirow{3}{*}{SFI}
            & CNN    & 931,140   & .883 $\pm$ .010 & .917 $\pm$ .005 & .883 $\pm$ .013 \\
            & MBC & 869,932   & .895 $\pm$ .009 & .918 $\pm$ .006 & .896 $\pm$ .007 \\
            & FNO    & 1,021,692 & \textbf{.906 $\pm$ .005} & \textbf{.921 $\pm$ .003} & \textbf{.899 $\pm$ .005} \\
        \bottomrule
    \end{tabular}
\end{table}

Additionally, we evaluate rate robustness across all five folds by perturbing test recordings by $95\%$, $90\%$, or a random $85$--$95\%$ factor (Table~\ref{tab:ablation}). To isolate \textit{sr} sensitivity from information loss due to downsampling, we retain only recordings whose occupied bandwidth remains safely representable after perturbation. Unaugmented SFI-FNO is strongly rate-sensitive, with a mean Macro-F1 change of $-.291\pm.074$, versus $-.025\pm.016$ for MAX-FNO. In contrast, rate augmentation reduces the mean change to $-.004\pm.005$, while maintaining Macro-F1 between .865 and .869 across perturbations.

\begin{table}[htbp]
    \centering
    \scriptsize
    \setlength{\tabcolsep}{3.1pt}
    \caption{Macro-F1 under \textit{sr} perturbation (mean $\pm$ std.).}
    \label{tab:ablation}
    \vspace{-0.1cm}
    \begin{tabular}{@{}lcccc@{}}
        \toprule
        \textbf{Model} & \textbf{$95\%$} & \textbf{$90\%$} & \textbf{Rand. ($85$-$95\%$)} & \textbf{Mean $\Delta$F1} \\
        \midrule
        SFI-FNO & $.703 \pm .031$ & $.555 \pm .024$ & $.563 \pm .025$ & $-.291 \pm .074$ \\
        SFI-FNO + aug. & $.865 \pm .007$ & \textbf{.869 $\pm$ .007} & \textbf{.869 $\pm$ .006} & $\mathbf{-.004 \pm .005}$ \\
        MAX-FNO & \textbf{.867 $\pm$ .011} & $.851 \pm .014$ & $.855 \pm .011$ & $-.025 \pm .016$ \\
        \bottomrule
    \end{tabular}
\end{table}


\begin{table}[t]
\centering
\scriptsize
\caption{Per-dataset performance of the SFI-FNO model, reported as mean $\pm$ standard deviation over five folds.}
\label{tab:sfi_fno_per_dataset}
\begin{tabularx}{\linewidth}{l *{3}{>{\centering\arraybackslash}X}}
\toprule
\textbf{Dataset} & \textbf{Accuracy} & \textbf{Balanced Acc.} & \textbf{Macro F1} \\
\midrule
Marine \cite{sayigh2016watkins} & $.933 \pm .012$ & $.942 \pm .006$ & $.903 \pm .011$ \\
Birds \cite{biodcase_tiny_2026_dataset} & $.675 \pm .030$ & $.675 \pm .030$ & $.673 \pm .029$ \\
Frogs \cite{akbal2023explainable} & $.996 \pm .004$ & $.991 \pm .004$ & $.991 \pm .004$ \\
Bats \cite{ndiaye2026bat_calls} & $.866 \pm .017$ & $.859 \pm .019$ & $.863 \pm .017$ \\
\bottomrule
\end{tabularx}
\end{table}

\vspace{-0.4cm}
\section{Discussion and Conclusions}
\vspace{-0.2cm}
First, it is important to note that reported performance in the literature varies across evaluation scopes. On cross-taxa benchmarks, for instance, AVES-Bio \cite{hagiwara2023aves} reports accuracies of $.879$ on Watkins and $.748$ on bats, while Perch~2.0 \cite{van2025perch} reaches respectively approximately $.86$--$.90$ and $.77$--$.82$. These values, as well as the higher accuracies reported by specialized Watkins-only systems \cite{lu2021detection, licciardi2024whalenet}, are not directly comparable with ours, as they evaluate dataset-specific classifiers; instead, we jointly classify $84$ classes spanning four taxonomic groups - Table \ref{tab:sfi_fno_per_dataset}. Therefore, our results should be interpreted in terms of robustness to heterogeneous taxa.

Secondly, fixed-rate resampling introduces a trade-off between spectral preservation and computational cost. Previous work on the Watkins dataset \cite{lu2021detection, cai2022parallel} resample recordings to $10$\,kHz, whereas WhaleNet \cite{licciardi2024whalenet} adopts $47.6$\,kHz after explicitly noting that a $10$\,kHz target would downsample a large part of the corpora. Similar fixed-rate assumptions remain common in recent bioacoustic learning systems, including methods for cross-species generalization such as \cite{nolasco2023learning, moummad2024regularized}. While such choices can be perfectly suitable for specific taxa sharing similar or predominantly low-frequency vocalizations, e.g., \cite{rauch2025birdmae}, they can become problematic in broader cross-taxa scenarios, or when datasets contain heterogeneous sampling rates due to differences in acquisition protocols or target frequency ranges, as in \cite{sayigh2016watkins}. Here, the frontend comparison is informative: increasing the \textit{sr} to the corpus-maximum improves all backbones, supporting that preserving the wide bandwidth is indeed beneficial. However, SFI consistently further improves over such resampling. Unlike MEL48 and MAX, SFI-FNO can exploit sample-rate-dependent shortcuts; however, the augmentation ablation shows that these are largely suppressed by rate augmentation, which reduces the mean perturbation-induced Macro-F1 change from $-.291\pm.074$ to $-.004\pm.005$ while preserving overall performance.

Finally, this distinction is computationally relevant. Downsampling to a moderate \textit{sr} reduces subsequent processing cost, but irreversibly discards frequency content above Nyquist limit; conversely, resampling to a high corpus-wide rate preserves the bandwidth, but increases the size of lower-rate recordings without introducing new information: this compromise is key in heterogeneous corpora. Recent work has indeed highlighted the influence of available frequency range on bioacoustic representations \cite{cauzinille2025species} and the importance of addressing computational costs in resource-constrained scenarios \cite{ciapponi2026enabling}. Thus, the proposed SFI frontend retains the actual bandwidth, providing a common representation without requiring \textit{sr} standardization.

To conclude, the proposed SFI-FNO framework combines native-rate spectral processing with global frequency-time modeling. The results provide a basis for extending this approach toward larger-scale, heterogeneous bioacoustic monitoring scenarios and foundation-model-like training approaches, such as unsupervised or semi-supervised learning.




\clearpage
{\small
\bibliographystyle{IEEEbib}
\bibliography{_REFS}
}

\end{document}